\documentstyle[11pt,epsfig]{article}
\newcommand{\beq}{\begin{equation}}
\newcommand{\eeq}{\end{equation}}
\newcommand{\beqa}{\begin{eqnarray}}
\newcommand{\eeqa}{\end{eqnarray}}
\newcommand{\beqan}{\begin{eqnarray*}}
\newcommand{\eeqan}{\end{eqnarray*}}
\newcommand{\ba}{\begin{array}}
\newcommand{\ea}{\end{array}}
\newcommand{\ben}{\begin{enumerate}}
\newcommand{\een}{\end{enumerate}}
\newcommand{\bfl}{\begin{flushleft}}
\newcommand{\efl}{\end{flushleft}}
\newcommand{\btab}{\begin{tabular}}
\newcommand{\etab}{\end{tabular}}
\newcommand{\bit}{\begin{itemize}}
\newcommand{\eit}{\end{itemize}}
\newcommand{\nl}{\nonumber \\}
\newcommand{\no}{\nonumber}
\newcommand{\ul}{\underline}

\newcommand{\ra}{\rightarrow}

\newcommand{\ve}{\varepsilon}
\newcommand{\vp}{\varphi}

\newcommand{\dg}{\dagger}

\newcommand{\wh}{\widehat}

\newcommand{\Ha}{{\cal H}}

\newcommand{\cL}{{\cal L}}

\newcommand{\dfrac}{\displaystyle \frac}

\newcommand{\toG}{\stackrel{G}{\to}}

\normalsize
\begin{document}
\begin{titlepage}
\begin{flushright}
UWThPh-1995-36\\
November 1995\\
\end{flushright}

\vspace{2.5cm}

\begin{center}
{\Large \bf Low--Energy QCD*}\\[40pt]
G. Ecker

\vspace{1cm}

Institut f\"ur Theoretische Physik, Universit\"at Wien\\
Boltzmanngasse 5, A--1090 Wien, Austria \\[10pt]

\vfill

\end{center}
\begin{abstract}
\noindent
After a brief introduction to chiral perturbation theory, the
effective field theory of the standard model at low energies,
two recent applications are reviewed: elastic pion--pion scattering
to two--loop accuracy and the complete renormalized pion--nucleon
Lagrangian to $O(p^3)$ in the chiral expansion.
\end{abstract}

\vfill
\begin{center}
To appear in the Proceedings of \\
International School of Nuclear Physics: Quarks in Hadrons and Nuclei\\
Erice, Sept. 19 -- 27, 1995\\
Progress in Particle and Nuclear Physics, Vol. 36
\end{center}
\vfill

\noindent * Work supported in part by FWF (Austria), Project No. P09505--PHY
and by HCM, EEC--Contract No. CHRX--CT920026 (EURODA$\Phi$NE)
\end{titlepage}

\renewcommand{\thesection}{\arabic{section}}
\renewcommand{\theequation}{\arabic{section}.\arabic{equation}}

%before each section
\setcounter{equation}{0}

\begin{center}
\section{THE STANDARD MODEL AT LOW ENERGIES}
\label{sec:SM}
\end{center}
At low energies ($E\ll 1~{\rm GeV}$), the simplicity of the QCD Lagrangian
is deceptive. There are no ``direct"
signs of quarks and gluons in the confinement regime. Instead, the
relevant degrees of freedom are first of all the hadrons that are
stable under the strong interactions: the pseudoscalar mesons and
the lowest--lying baryons. At a second stage, meson and baryon resonances
can be included.

In principle, we are told to integrate out the fundamental degrees
of freedom (quarks and gluons) to arrive at a field theory of the observed
hadronic fields. In the confinement regime, this procedure is
under theoretical control only for the chiral anomaly (Wess and Zumino,
1971). In practice therefore, one uses the symmetries of QCD and of the
standard model
in general to arrive at an effective field theory at low energies
(Weinberg, 1979) called chiral perturbation theory (CHPT) (Gasser and
Leutwyler, 1984, 1985; Leutwyler, 1994a). The crucial role in the
construction of this
effective field theory is played by the spontaneously broken chiral
symmetry with the pseudoscalar mesons as corresponding (pseudo-) Goldstone
bosons. Referring to recent reviews (Bernard et al., 1995; Ecker, 1995b;
Leutwyler, 1994b; Pich, 1995; de Rafael, 1995) for
a more extensive coverage of CHPT, I list
here only a few salient features.
\bit
\item As is the case for effective field theories with spontaneously
broken symmetries in general, CHPT is a non--renormalizable quantum
field theory.
\item As required by unitarity, a consistent low--energy expansion entails
a loop expansion. Since the loop amplitudes are in general divergent, the
theory has to be regularized and renormalized.
\item Once this has been achieved, CHPT incorporates all the usual properties
of a bona fide quantum field theory, at least in the perturbative sense:
unitarity, analyticity, crossing, \dots
\item The symmetries of the standard model are manifest by construction.
\item There is no double--counting in CHPT: only hadronic fields,
but neither quarks nor gluons appear in the chiral Lagrangian.
\item All the short--distance structure is encoded in certain
coupling constants, the so--called low--energy constants (LECs). In
pure CHPT with only pseudoscalar mesons and low--lying
baryons, even the resonances are included among the short--distance effects.
More generally, the LECs describe the influence of all degrees of freedom
not explicitly contained in the effective chiral Lagrangian.
\item CHPT is not a special model for low--energy QCD like, e.g.,
the Nambu--Jona-Lasinio model (Nambu and Jona-Lasinio, 1961), but a
quantum field theory framework to construct the most general solution
of the Ward identities of QCD (the standard model in general).
\eit

There is a price to pay for this generality. Since the LECs parametrize
the solutions of the Ward identities, they are by definition not constrained
by the symmetries. Additional input is needed to make CHPT predictive,
especially in higher orders of the chiral expansion. This information
(for a recent review, see Ecker, 1995a) comes either from experiment or
from additional theoretical input
(resonance saturation, QCD sum rules, large--$N_c$ expansion, lattice
QCD, Nambu--Jona-Lasinio type models, skyrmions, \dots).

%before each section
\setcounter{equation}{0}

\vspace*{.5cm}

\begin{center}
\section{EFFECTIVE CHIRAL LAGRANGIAN}
\label{sec:EFT}
\end{center}
In the real world, there is no chiral symmetry. Even in the limit of
vanishing quark masses, all theoretical and phenomenological evidence
indicates that the chiral group  $G=SU(N_f)_L \times SU(N_f)_R$
(for $N_f$=2 or 2 massless quarks) is spontaneously broken to the
diagonal (vectorial) subgroup $SU(N_f)_V$. There is a standard procedure
(Coleman et al., 1969; Callan et al., 1969) how to realize a
spontaneously broken symmetry on quantum fields. In the special case
of chiral symmetry with its parity transformation, the Goldstone
fields $\vp$ can be collected in a unitary matrix field $U(\vp)$
transforming as
\beq
U(\vp) \toG g_R U(\vp) g_L^{-1} ~,\qquad
(g_L,g_R) \in G \label{Utr}
\eeq
under chiral transformations.

The chiral symmetry is in addition broken explicitly by
non--vanishing quark masses and, if the weak interactions are included,
through the handedness of the weak interactions. Restricting the attention
first to the strong, electromagnetic and semileptonic weak interactions,
the most convenient way to introduce the explicit chiral symmetry breaking
is via the introduction of external scalar ($s$), pseudoscalar ($p$),
vector ($v$) and axial--vector ($a$) fields (Gasser and
Leutwyler, 1984, 1985) that contain both the quark masses and external
photons and $W$ bosons.

Although this framework is sufficient for the applications
I am going to discuss in this talk, let me emphasize for completeness
that the formalism must be extended if one wants to include ``internal"
photons and $W$ bosons. Turning to the non--leptonic weak interactions,
one first has to integrate out the $W$ boson together with the
heavy quarks to arrive at an effective Hamiltonian still at the
fundamental quark level that transforms under chiral
transformations as
\beq
\Ha_{\rm eff}^{\Delta S =1} \sim (8_L,1_R) + (27_L,1_R)
\label{eq:nldecomp}
\eeq
in the $\Delta S =1$ sector. This effective Hamiltonian is then realized
by an effective chiral Lagrangian at the level of mesons and
baryons that must, of course, have the same transformation property
(\ref{eq:nldecomp}).

The situation is again different for virtual photons relevant for the
treatment of electromagnetic corrections. Virtual photons cannot be
integrated out to produce a local Hamiltonian at the quark level. Instead,
one has to introduce the photon as a dynamical field at the
hadronic level. In addition, one must include the general chiral Lagrangian
of $O(e^2)$ that transforms as the product of two electromagnetic currents.
The mesonic Lagrangian of $O(e^2 p^2)$ has only recently been constructed
(Urech, 1995; Neufeld and Rupertsberger, 1995).

CHPT is based on a two--fold expansion. As a
low--energy effective field theory, it is an expansion in small
momenta. On the other hand, it is also an expansion in the quark masses
$m_q$ around the chiral limit. In full generality,
the effective chiral Lagrangian is of the form
\beq
\cL_{\rm eff} = \sum_{i,j} \cL_{ij}~, \qquad \qquad
\cL_{ij} = O(p^i m^j_q)~. \label{eq:gexp}
\eeq
The two expansions become related by expressing the pseudoscalar meson
masses in terms of the quark masses. If the quark condensate is
non--vanishing in the chiral limit, the squares of the meson masses
start out linear in $m_q$ [cf. Eq.~(\ref{eq:mpi2})].
Assuming the linear terms to give the dominant contributions to the
meson masses, one arrives at the standard chiral counting (Gasser and
Leutwyler, 1984,1985) with $m_q = O(p^2)$  and
\beq
\cL_{\rm eff} = \sum_d \cL_d~, \qquad \qquad
\cL_d = \sum_{i + 2j = d} \cL_{ij}~.
\eeq

An alternative way to organize the chiral expansion accounts for the
possibility that the quark condensate might be much smaller than
usually assumed. In that case, the leading--order contributions according
to the usual counting would not necessarily give the dominant contributions
to the meson masses. Although there are several arguments in favour of the
standard counting, among them the validity of the Gell-Mann--Okubo
mass formula for the pseudoscalar meson masses, the alternative approach
of Generalized CHPT (see Knecht and Stern, 1995a for a recent review) is
still a logical possibility.
In the generalized picture, more terms appear at a
given order that are relegated to higher orders in the standard counting.
Obviously, this procedure increases the number of LECs
at any given order. It should be kept in mind, however, that the effective
chiral Lagrangian of the standard model is the same in the standard
and in the generalized approach.
In this talk, I will strictly adhere to the standard procedure, but
I will briefly come back to Generalized CHPT in the discussion of
$\pi\pi$ scattering.

The effective chiral Lagrangian of the standard model is shown
schematically in Table \ref{tab:EFTSM}. The subscripts of the different
parts of this Lagrangian denote the chiral dimension according to
the standard counting and the numbers in brackets indicate the appropriate
number of LECs. The notation even/odd refers to the
mesonic Lagrangians without/with an $\varepsilon$ tensor (even/odd intrinsic
parity). I have grouped together those pieces of the Lagrangian that
have the same chiral order as a corresponding loop amplitude ($L$ =
0, 1, 2). The Lagrangians ${\cal L}_n^{\Delta S=1}$ and
${\cal L}_n^\gamma$ describe non--leptonic weak
interactions and virtual photons, respectively, but I have only included
the purely mesonic parts. In fact, in the meson--baryon sector only
the pion--nucleon Lagrangian is included, i.e. $N_f=2$. On the other
hand, the number of LECs in the purely mesonic Lagrangians are given
for $N_f=3$. As already emphasized in the beginning, the theory has
to be renormalized once one reaches a chiral order where loop diagrams
must be included. The parts of the effective chiral Lagrangian that
have been completely renormalized are underlined in Table
\ref{tab:EFTSM}.

\renewcommand{\arraystretch}{1.1}
\begin{table}
\caption{The effective chiral Lagrangian of the standard model}
\label{tab:EFTSM}
$$
\begin{tabular}{lc|c}
\hspace{2cm} ${\cal L}_{\rm chiral \,\, dimension}$ ~($\#$ of LECs)
& \hspace{1.5cm} & loop order \\[10pt]
\hline
& & \\[10pt]
${\cal L}_2(2)$~+~${\cal L}_4^{\rm odd}(0)$~+~${\cal L}_2^{\Delta S=1}(2)$
{}~+~${\cal L}_0^\gamma(1)$  & \hspace{1.5cm} & $L=0$ \\[10pt]
{}~+~${\cal L}_1^{\pi N}(1)$~+~${\cal L}_2^{\pi N}(7)$~+~\dots & & \\[20pt]
{}~+~$\ul{{\cal L}_4^{\rm even}(10)}$~+~$\ul{{\cal L}_6^{\rm odd}(32)}$
{}~+~$\ul{{\cal L}_4^{\Delta S=1}(22,{\rm octet})}$~+
{}~$\ul{{\cal L}_2^\gamma(14)}$ & &  $L=1$ \\[10pt]
{}~+~$\ul{{\cal L}_3^{\pi N}(24)}$~+~${\cal L}_4^{\pi N}(?)$~+~\dots & &
\\[20pt]
{}~+~${\cal L}_6^{\rm even}(111)$~+~\dots & & $L=2$ \\[12pt] \hline
\end{tabular}
$$
\end{table}

The Table shows the dramatic increase of the number of
LECs with the chiral order. It is clear that one will never
be able to measure all 111 LECs (Fearing and Scherer, 1994) in the strong
meson Lagrangian of $O(p^6)$. As I will try to demonstrate, one can
make meaningful predictions to $O(p^6)$ nevertheless.

%before each section
\setcounter{equation}{0}

\vspace*{.5cm}

\begin{center}
\section{ELASTIC PION--PION SCATTERING}
\label{sec:pipi}
\end{center}
Pion--pion scattering is a fundamental process for testing CHPT
that involves only the pseudo--Goldstone bosons of chiral
$SU(2)$. In the limit of isospin conservation ($m_u=m_d$),
the scattering amplitude for
\beq
\pi^a(p_a) + \pi^b(p_b) \to \pi^c(p_c) + \pi^d(p_d)
\eeq
is determined by a single scalar function $A(s,t,u)$ defined by the
isospin decomposition
\beqa
T_{ab,cd} &=& \delta_{ab}\delta_{cd} A(s,t,u) + \delta_{ac}\delta_{bd}
A(t,s,u) + \delta_{ad}\delta_{bc} A(u,t,s) \nl
A(s,t,u) &=& A(s,u,t)
\eeqa
in terms of the usual Mandelstam variables $s,t,u$. The amplitudes
$T^I(s,t)$ of definite isospin $(I = 0,1,2)$ in
the $s$--channel are decomposed into partial waves ($\theta$ is the
center--of--mass scattering angle):
\beq
T^I(s,t)=32\pi\sum_{l=0}^{\infty}(2l+1)P_l(\cos{\theta})t_l^I (s)~.
\eeq
Unitarity implies that in the elastic region $4M_\pi^2 \leq s
\leq 16M_\pi^2$ the partial--wave amplitudes $t_l^I$ can be described
by real phase shifts $\delta_l^I$.
The behaviour of the partial waves  near threshold is of the form
\beq
\Re e\;t_l^I(s)=q^{2l}\{a_l^I +q^2 b_l^I +O(q^4)\}~,
\eeq
with $q$ the center--of--mass momentum.
The quantities $a_l^I$ and $b_l^I$ are called
scattering lengths and slope parameters, respectively.

At lowest order in the chiral expansion, $O(p^2)$, the scattering amplitude
is given by the current algebra result (Weinberg, 1966)
\beq
A_2(s,t,u)=\dfrac{s-M_\pi^2}{F_\pi^2}~, \label{eq:A2}
\eeq
leading in particular to an $I=0$ S--wave scattering length
$a_0^0=0.16$. Near threshold, the chiral expansion for
$SU(2)$ is expected to converge rapidly because the
natural expansion parameter is of the order
\beq
\dfrac{4 M^2_\pi}{16 \pi^2 F^2_\pi} = 0.06~.
\eeq
However, CHPT produces also singularities in the quark mass expansion
(the so--called chiral logarithms) that may enhance this value.
For an $L$--loop amplitude, the chiral logarithms appear
in a general amplitude as $(\ln{\dfrac{p^2}{\mu^2}})^n$ with $n\le L$.
Here, $p$ is a generic momentum and the dependence on the arbitrary
scale $\mu$ is compensated by the scale dependence of the appropriate
LECs in the amplitude.

The scattering amplitude of $O(p^4)$ (Gasser and Leutwyler, 1983, 1984)
has the general structure
\beqa
F_\pi^4 A_4(s,t,u) =& c_1 M_\pi^4 + c_2 M_\pi^2 s + c_3 s^2 +
c_4 (t-u)^2 \nl
&+ F_1(s) + G_1(s,t) + G_1(s,u)~.
\eeqa
The functions $F_1, G_1$ are one--loop functions and the coefficients
$c_1,\dots,c_4$ of the general crossing symmetric polynomial of $O(p^4)$
are given in terms of the appropriate LECs (Gasser and Leutwyler,
1983, 1984) $l_i^r(\mu) ~(i=1,\dots,4)$ and of $\ln{M_\pi^2/\mu^2}$.
Since four different LECs appear at this order, it is not surprising that
chiral symmetry does not put any further constraints on the $c_i$. With the
phenomenological values of the LECs $l_i^r(\mu)$, Gasser and Leutwyler
(1983, 1984) obtained $a_0^0=0.20 \pm 0.01$ to be compared with the
value $a_0^0=0.26 \pm 0.05$ extracted from experiment (Nagels et al.,
1979). One observes first of all quite a sizeable correction of $O(p^4)$
which can be attributed to a large extent to chiral logs and, secondly,
a still bigger experimental value, albeit with a large error.

The value of $a_0^0$ has some bearing on the mechanism of spontaneous
chiral symmetry breaking. If the pseudoscalar meson masses are not
dominated by the lowest--order contributions proportional to the quark
condensate [cf. Eq.~(\ref{eq:mpi2})], the standard quark mass ratios
could be significantly
modified. This is precisely the option that Generalized CHPT proposes
to keep in mind. In the generalized approach, the scattering lengths can
not be absolutely predicted at $O(p^2)$, but they
depend in addition on the quark mass ratio $r=2 m_s/(m_u+m_d)$
(Stern et al., 1993; Knecht et al., 1995b). Taking
the experimental mean value for $a_0^0$ at face value would
decrease the quark mass ratio $r$ from its generally accepted value 26 to about
10.

After a period of rather little activity on the experimental
side, there are now good prospects for more precise
data on $\pi\pi$ scattering in the near future. Most promising are the
forthcoming data
from $K_{e4}$ decays at the $\Phi$--factory DA$\Phi$NE in Frascati which
are expected to reduce the experimental uncertainty of $a_0^0$ to some
$5\%$ in $10^7 s$ running time (Baillargeon and Franzini, 1995).
These experimental prospects have prompted two groups to attack the
scattering amplitude of $O(p^6)$ (Knecht et al., 1995c; Bijnens et al.,
1995). Knecht et al. have used dispersion relations to calculate the
analytically non--trivial part of the amplitude and they have fixed
some of the subtraction constants using $\pi\pi$ scattering data at
higher energies. This approach does not yield the chiral logs which,
although contributing only to the polynomial
part of the amplitude, make an important numerical contribution
(Colangelo, 1995; Bijnens et al., 1995). The standard field theory
calculation of CHPT produces all those terms, but one has to calculate
quite a few diagrams with $L=0,1$ and 2 loops to get the final amplitude.
The two groups completely agree on the absorptive parts
which are contained in the functions $F_2$,
$G_2$ in the general decomposition
\beqa
F_\pi^6 A_6(s,t,u) =& d_1 M_\pi^6 + d_2 M_\pi^4 s + d_3 M_\pi^2 s^2 +
d_4 M_\pi^2 (t-u)^2 + d_5 s^3 + d_6 s (t-u)^2 \no \\*
&+ F_2(s) + G_2(s,t) + G_2(s,u)~.
\eeqa
The coefficients $d_1,\dots,d_6$ in the general crossing symmetric
polynomial of $O(p^6)$ depend on the LECs $l_i^r(\mu)$ ($i=1,
\dots,4$) of $O(p^4)$, on the chiral logs $(\ln{M_\pi^2/\mu^2})^n ~(n=1,2)$
and on six combinations of the LECs of $O(p^6)$ (Fearing
and Scherer, 1994). Again, chiral symmetry does not constrain these
six combinations. However, both chiral dimensional analysis and saturation
by resonance exchange, which is known to work very well at $O(p^4)$
(Ecker et al., 1989a, 1989b), suggest values for these LECs
that do not affect the threshold parameters in a dramatic way, especially
not for the $S$--waves. The coefficients $d_i$ are dominated
on the one hand by the LECs $l_i^r(\mu)$ of $O(p^4)$ and by the chiral
logs. Instead of ascribing a theoretical error to
$a_0^0$ (which is dominated by the errors of the
$l_i^r(\mu)$), I compare the predicted values to $O(p^n)$ for
$n=2,4$ and 6 (Bijnens et al., 1995) to three significant digits (using
the ``old" value $F_\pi=93.2 ~{\rm MeV}$ and the charged pion mass)~:
$$
\renewcommand{\arraystretch}{1.4}
\begin{tabular}{|c|c|c|c|} \hline
$n$ & 2 & 4 & 6 \\ \hline
$a_0^0$ & $0.156$ & $0.201$ & $0.217$  \\ \hline
\end{tabular}
$$

The main conclusion concerning $a_0^0$ is that the correction of
$O(p^6)$ is reasonably small and under theoretical control making
a value as high as $0.26$ practically impossible to accommodate within
standard CHPT. Therefore, the
forthcoming $\pi\pi$ data can be expected to either corroborate
this prediction with significant precision or to shed serious
doubts on the assumed mechanism of spontaneous chiral symmetry breaking
through the quark condensate.

For a discussion of other threshold parameters and of
the phase shifts themselves, I refer to Bijnens et al. (1995) for the
standard CHPT calculation and to Knecht et al. (1995c) for the dispersion
theory analysis.

At the level of precision we have reached with the $O(p^6)$ calculation,
one may wonder about the size of electromagnetic and isospin violating
corrections. Neglecting the tiny $\pi^0-\eta$ mixing angle, the charged
and neutral pion masses are equal to $O(p^2)$ without assuming
$m_u=m_d$:
\beq
M^2_{\pi^+}=M^2_{\pi^0}=(m_u+m_d) B ~,
\label{eq:mpi2}
\eeq
where $B$ is proportional to the quark condensate. The mass difference
between charged and neutral pions is almost entirely an effect of
$O(e^2 p^0)$ and it is determined by the Lagrangian ${\cal L}_0^\gamma(1)$
in Table 1. This effect is obviously non--negligible also for $\pi\pi$
scattering as can be seen, for instance, from the lowest--order expression
for the $I=0 ~S$--wave scattering length evaluated with either the
charged or the neutral pion mass:
\beq
a^0_0 = \dfrac{7 M^2_\pi}{32 \pi F^2_\pi} = \left\{\ba{ll}
0.156  &~\mbox{with} ~~M^2_{\pi^+} \\
0.146  &~\mbox{with} ~~M^2_{\pi^0}
\ea \right.~. \label{eq:a00+0}
\eeq
Clearly, the difference is comparable to the chiral correction of $O(p^6)$.
The pion decay constant is also affected by radiative corrections, which
have been estimated (Holstein, 1990; Review of Particle Properties,
1994) to move $F_\pi$ down from 93.2 to 92.4 MeV. This decrease of $F_\pi$
increases the $O(p^6)$ prediction for $a_0^0$ from 0.217 to 0.222
(Bijnens et al., 1995).

The question is then what other effects of $O(e^2 p^0)$ appear in
the $\pi\pi$ scattering amplitude. To answer this question, let us
restrict the Lagrangian ${\cal L}_0^\gamma(1)$ to $N_f=2$ and
expand it in pion fields, using for convenience the so--called
$\sigma$--parametrization for $U$:
\beq
{\cal L}_0^\gamma(1)({\rm pions})=e^2 C \langle Q U Q U^\dg\rangle
({\rm pions})= - \dfrac{2 e^2 C}{F^2}\pi^+\pi^-~,
\eeq
where $Q$ is the quark charge matrix, $C$ is the unique LEC of $O(e^2 p^0)$
and $<\dots>$ stands for the trace in two--dimensional flavour space.
The conclusion is that there are no terms of $O(\pi^n)$ for $n>2$. In
other words, the Lagrangian ${\cal L}_0^\gamma(1)$ contributes only
to the $\pi^+ - \pi^0$ mass difference, but not to the scattering
amplitude itself. To leading $O(e^2 p^0)$ therefore, electromagnetic
corrections appear only in the kinematics and can easily be accounted for.
The leading
non--trivial electromagnetic effects for $\pi\pi$ scattering occur at
$O(e^2 p^2)$ and they are under investigation (J. Gasser, private
communication). They can be expected to be quite a bit smaller
than suggested by Eq.~(\ref{eq:a00+0}).

%before each section
\setcounter{equation}{0}

\vspace*{.5cm}

\begin{center}
\section{PION--NUCLEON LAGRANGIAN TO $O(p^3)$}
\label{sec:piN}
\end{center}
We are looking for a systematic low--energy expansion of the pion--nucleon
Lagrangian for single--nucleon processes, i.e. for processes of the type
$\pi N \ra \pi \ldots \pi N$, $\gamma N \ra \pi \ldots \pi N$,
$\gamma^* N \ra \pi
\ldots \pi N$ (including nucleon form factors), $W^* N \ra \pi \ldots \pi N$.
There is an obvious problem with chiral counting here: in contrast to
pseudoscalar mesons, the nucleon four--momenta can never be ``soft" because
the nucleon mass does not vanish in the chiral limit. Although the problem
can be handled at the Lagrangian level (Gasser et al., 1988;
Krause, 1990), it reappears once one goes beyond tree level. The loop
expansion and the derivative expansion do not coincide
like in the meson sector. The culprit is again the nucleon mass that enters
loop amplitudes through the nucleon propagators. In the original relativistic
formulation (Gasser et al., 1988), amplitudes of a given chiral order
receive contributions from any number of loops.

A comparison between the nucleon mass and the chiral expansion scale
$4 \pi F_\pi$ suggests a simultaneous expansion in
$$
\dfrac{\vec{p}}{4\pi F} \qquad \mbox{and} \qquad \dfrac{\vec{p}}{m}
$$
where $\vec{p}$ is a small three--momentum.
On the other hand, there is a crucial difference between $F\simeq F_\pi$
and $m\simeq m_N$: whereas $F$ appears only in vertices, the
nucleon mass enters a generic diagram via the nucleon propagator. The
idea of Heavy Baryon CHPT (HBCHPT) (Jenkins and Manohar, 1991,
1992) is precisely to transfer $m$ from the propagators to some vertices.
The method can be interpreted as a clever choice of fermionic variables
(Mannel et al., 1992) in the generating functional of Green functions
(Gasser et al., 1988)
\beq
e^{iZ[j,\eta,\bar\eta]} =  \int [du d\Psi d \bar \Psi]
\exp [i\{ S_M + S_{\pi N} + \int d^4 x (\bar \eta \Psi + \bar \Psi \eta)\}]~.
\label{eq:ZpiN}
\eeq
Here, $S_M + S_{\pi N}$ is the combined pion--nucleon action in the
relativistic framework, $\Psi$ is the nucleon field, $\eta$ is a
fermionic source and $j$ stands for the previously introduced external
fields ($v,a,s,p$).

In terms of velocity dependent fields $N_v,H_v$ defined as (Georgi, 1990)
\beqa
N_v(x) &=& \exp[i m v \cdot x] P_v^+ \Psi(x) \label{eq:vdf} \\
H_v(x) &=& \exp[i m v \cdot x] P_v^- \Psi(x) \no \\
P_v^\pm &=& \frac{1}{2} (1 \pm \not\!v)~, \qquad v^2 = 1 ~, \no
\eeqa
with a time--like unit four--vector $v$, the pion--nucleon action
$S_{\pi N}$ takes the form
\beqa
S_{\pi N} &=& \int d^4 x \{ \bar N_v A N_v + \bar H_v B N_v +
\bar N_v \gamma^0 B^\dg \gamma^0 H_v - \bar H_v C H_v\} \label{eq:SpiN} \\
I &=& I_{(1)}+ I_{(2)} + I_{(3)} + \ldots~, \qquad I=A,B,C ~. \no
\eeqa
The operators $A_{(n)}$, $B_{(n)}$, $C_{(n)}$ are the corresponding projections
of the relativistic pion--nucleon Lagrangians $\cL_{\pi N}^{(n)}$.
Rewriting also the source term in (\ref{eq:ZpiN}) in terms of
$N_v, H_v$,
one can integrate out the ``heavy'' components $H_v$
 to obtain a non--local action in the
fields $N_v$ (Bernard et al., 1992). Expanding this
non--local action in a power series
in $1/m$, one obtains a Lorentz--covariant chiral expansion
for the Lagrangian
\beq
\wh \cL_{\pi N}(N_v;v)=\wh \cL_{\pi N}^{(1)}+\wh \cL_{\pi N}^{(2)}+
\wh \cL_{\pi N}^{(3)}+\dots~, \label{eq:pinexp}
\eeq
which depends of course on the arbitrary
four--vector $v$. Specializing to the nucleon rest frame (either in the
initial or in the final state) with $v=(1,0,0,0)$, (\ref{eq:pinexp})
amounts to a
non--relativistic expansion for the $\pi N$ Lagrangian. In this Lagrangian,
the nucleon mass $m$ appears only in vertices, but not in the propagator of
the transformed nucleon field $N_v$.

A given Lorentz covariant Lagrangian for the field $N_v$
will in general not be Lorentz invariant because it depends on the
arbitrary four--vector $v$. To guarantee Lorentz invariance, two
procedures are possible. One can either impose reparametrization
invariance (Luke and Manohar, 1992) on the Lagrangian (\ref{eq:pinexp})
a posteriori
or one can start directly from the fully relativistic Lagrangian which is
Lorentz invariant by construction. The second approach has
advantages especially in higher orders of the chiral expansion and it
implies of course reparametrization invariance. This is the
approach I am going to follow here.

The relativistic pion--nucleon Lagrangian of lowest order (Gasser
et al., 1988),
\beq
\cL_{\pi N}^{(1)} = \bar \Psi \left(i \not\!\nabla - m + \frac{g_A}{2}
\not\!u \gamma_5\right)\Psi, \label{LMB1}
\eeq
leads directly to the corresponding ``non--relativistic" Lagrangian of
$O(p)$:
\beq
\wh \cL_{\pi N}^{(1)} = \bar N_v(iv \cdot \nabla + g_A S \cdot u) N_v~.
\label{piN1}
\eeq
Here,  $m$ and $g_A$ are the nucleon mass and the axial--vector coupling
constant in the chiral limit, $\nabla$ is a covariant derivative that
includes in particular the photon field, $u^\mu$ is the vielbein field
related to the matrix field $U$ and $S^\mu = i \gamma_5 \sigma^{\mu\nu}
v_\nu/2$ is the spin matrix, the only remnant of Dirac matrices in
the Lagrangian $\wh \cL_{\pi N}$.

At the next chiral order, $O(p^2)$, the Lagrangian $\wh \cL_{\pi N}^{(2)}$
consists of two pieces (Bernard et al., 1992). There is first a piece
that is due to the
expansion in $1/m$ with completely determined coefficients and there
is in addition the non--relativistic reduction of the relativistic
Lagrangian $\cL_{\pi N}^{(2)}$. After a suitable field transformation
of the nucleon field $N_v$, the
Lagrangian assumes its most compact form (Ecker and Moj\v zi\v s, 1995c)
\beqa
\wh \cL_{\pi N}^{(2)} &=& \bar N_v\left( - \frac{1}{2m} (\nabla \cdot
\nabla + ig_A \{S \cdot \nabla, v \cdot u\}) \right. \no \\
&& \mbox{} + \frac{a_1}{m} \langle u \cdot u\rangle +
\frac{a_2}{m} \langle (v \cdot u)^2\rangle +
\frac{a_3}{m} \langle \chi_+\rangle +
\frac{a_4}{m} \left( \chi_+ - \frac{1}{2} \langle \chi_+\rangle \right)
\no \\
&& \left. \mbox{} + \frac{1}{m} \ve^{\mu\nu\rho\sigma} v_\rho S_\sigma
[i a_5 u_\mu u_\nu + a_6 f_{+\mu\nu} + a_7 v_{\mu\nu}^{(s)}]\right) N_v~,
\label{piN2}
\eeqa
where $\chi_+$ contains the quark mass matrix and
the tensor fields $f_{+\mu\nu}$, $v_{\mu\nu}^{(s)}$
are the isovector and isoscalar parts of the external gauge fields
including the electromagnetic field.
The LECs $a_i$ ($i=1,\dots,7$) are dimensionless and expected to be
of $O(1)$ according to naive chiral dimensional analysis. They are
in fact all known phenomenologically and I refer to Bernard et al. (1995)
for an up-to-date review. For future purposes, let me single out
two of them that are related to the nucleon magnetic moments in the
chiral limit:
\beqa
\label{nmm}
a_6 &=& \frac{\mu_v}{4} = \frac{1}{4} (\mu_p - \mu_n) \no \\
a_7 &=& \frac{\mu_s}{2} = \frac{1}{2} (\mu_p + \mu_n)~.
\eeqa

The first two terms in the Lagrangian (\ref{piN2}) illustrate
the difference between Lorentz covariance and invariance. The latter
fixes the coefficients uniquely although covariance alone would
seem to allow arbitrary coefficients. That these coefficients
cannot be arbitrary becomes obvious when one realizes that the first
term governs the Thomson limit for nucleon Compton scattering and the
second one is responsible for the $O(p^2)$ contribution for pion
photoproduction on nucleons at threshold. Of course, both amplitudes are
completely determined by the nucleon charge and by $g_A$.

Chiral power counting (Weinberg, 1990) for the chiral dimension
$D$ of a generic CHPT diagram with $N_d^M$
mesonic vertices and $N_d^{MB}$ pion--nucleon vertices of $O(p^d)$,
\beq
D = 2L + 1 + \sum_d (d-2) N_d^M + \sum_d (d-1) N_d^{MB} \geq 2L+1~,
\label{eq:DMB2}
\eeq
tells us that loop diagrams enter at $O(p^3)$. As is to be expected
in a non--renormalizable quantum field theory, those loop diagrams
are in general divergent and must be regularized. Consequently, the
theory has to be renormalized to give results independent of the
regularization procedure. The structure of the divergences
(the divergence functional) can be calculated in closed form (Ecker,
1994) leading to scale--dependent LECs of $O(p^3)$. After applying
again suitable field transformations, the complete $\pi N$ Lagrangian
of $O(p^3)$ is found to be (Ecker and Moj\v zi\v s, 1995c)
\beqa
\label{piN3}
\wh \cL_{\pi N}^{(3)} &=& \bar N_v \left( \frac{g_A}{8 m^2}
[ \nabla_\mu,[\nabla^\mu,S \cdot u]] + \frac{1}{2m^2} \left[
\left\{ i \left( a_5 - \frac{1-3g^2_A}{8}\right) u_\mu u_\nu
 \right. \right. \right. \no \\
&& \mbox{} + \left. \left( a_6 - \frac{1}{8}\right) f_{+\mu\nu} +
\left( a_7 - \frac{1}{4}\right) v_{\mu\nu}^{(s)} \right\}
\ve^{\mu\nu\rho\sigma} S_\sigma i \nabla_\rho + \frac{g_A}{2}
S \cdot \nabla u \cdot \nabla \no \\
&& \mbox{} - \left. \frac{g^2_A}{8} \{v \cdot u,u_\mu\}
\ve^{\mu\nu\rho\sigma} v_\rho S_\sigma \nabla_\nu - \frac{ig_A}{16}
\ve^{\mu\nu\rho\sigma} f_{-\mu\nu} v_\rho \nabla_\sigma + {\rm h.c.}\right]
\no \\
&& \mbox{} + \left.\frac{1}{(4\pi F)^2} \sum_{i=1}^{24} b_i O_i \right)
N_v~.
\eeqa
Although quite a bit more involved, this Lagrangian has the same structure
as (\ref{piN2}). There is a piece with coefficients completely fixed
in terms of LECs of $O(p)$ and $O(p^2)$. The second part has 24 new
LECs $b_i$. The associated field monomials $O_i$ can be
found in Ecker and Moj\v zi\v s (1995c). It is this second part that
is needed to absorb the divergences of the one--loop functional. The
splitting of the $b_i$ into divergent and finite parts introduces a
scale dependence of the finite, measurable LECs $b_i^r(\mu)$. This
scale dependence is governed by $\beta$--functions that are determined
by the divergence functional (Ecker, 1994). Adding the finite part of
the one--loop functional, one arrives at the total generating
functional of Green functions in the pion--nucleon system to $O(p^3)$:
\beq
Z = Z_1(g_A) + Z_2(a_i;g_A,m) +
 Z_{3,{\rm finite}}^{L=1}(g_A;\mu) + Z_3^{\rm tree}(b_i^r(\mu);
a_i,g_A,m)~. \label{eq:Z}
\eeq
The functionals $Z_1$, $Z_2$ are tree--level functionals, whereas
the functional of $O(p^3)$ consists of both a loop and a tree--level
part. The sum of those two and therefore the complete functional is
independent of the arbitrary scale $\mu$.

The functional (\ref{eq:Z}) contains the complete low--energy structure
of the $\pi N$ system to $O(p^3)$. In order to extract physical amplitudes
from this functional, one has to calculate the appropriate
one--loop amplitudes contained in $Z_3^{L=1}$. This has already been
done for many processes of interest and I refer especially to the review
of Bernard et al. (1995) for an extensive coverage of the phenomenological
applications. Here, I want to consider an illustrative example of the
class of amplitudes (or Green functions more generally) that are
insensitive to the LECs $b_i$ of $O(p^3)$. This class of amplitudes
is of course of special interest for comparison with experiment because
one does not need to know anything about the actual values of the
renormalized LECs $b_i^r$. This is welcome because we are far from
having very reliable information on most of these LECs.

This class of amplitudes can still be divided into two groups. In
the first group, loop amplitudes do contribute, which are then necessarily
finite because there are no available counterterms that
could absorb the divergences. A well--known example is neutral pion
photoproduction at threshold, $\gamma N \to \pi^0 N$, where it has
been found only
relatively recently (Bernard et al., 1991) that there is a sizeable
loop contribution of $O(p^3)$. In this talk I want to discuss
an example of the second class where there are neither loop nor
counterterm contributions at $O(p^3)$. The only possible other
contribution at this
order must then come from the terms with fixed coefficients in (\ref{piN3}).

As an example, consider nucleon Compton scattering at small photon
energies in the forward direction. In a gauge where the
polarization vectors have vanishing time components, the forward
scattering amplitude has the form
\beq
T = c_0 \, \vec{\ve}\, ' \cdot \vec{\ve} + i \, c_1 \, \delta
\, \vec{\sigma} \cdot (
\vec{\ve}\, ' \times \vec{\ve} \, ) +  O(\delta^2) ~ ,
\label{Comp}
\eeq
where $\delta=E_\gamma/m$. The coefficient $c_0$ determining the leading
spin--independent amplitude of $O(p^2)$ is given by the Thomson limit
\beq
c_0 = -\dfrac{q_N e^2}{4\pi m}
\eeq
with $q_N$ the nucleon charge. As previously mentioned, this coefficient
is directly given by the first term in the Lagrangian (\ref{piN2}).

\begin{figure}
\centerline{\epsfig{file=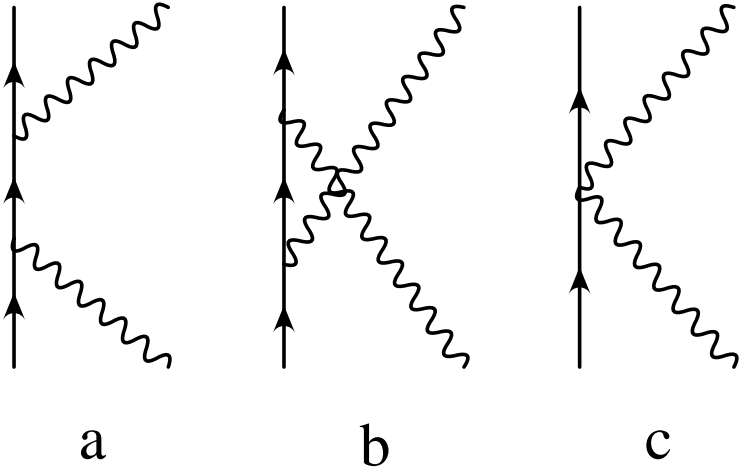,height=7cm}}
\caption{Tree diagrams for nucleon Compton scattering that are
responsible for the leading coefficients $c_0$, $c_1$ of the forward
scattering amplitude at low photon energies.}\label{compton}
\end{figure}

How does HBCHPT account for the leading spin--dependent Compton
amplitude in terms of $c_1$? It is not very difficult to convince
oneself that there is indeed neither a loop contribution nor a
contribution proportional to the LECs $b_i$. The relevant diagrams
are shown in Fig. \ref{compton} where the vertices in diagrams a,b
are due to the Lagrangian (\ref{piN2}), while the seagull vertex
of diagram c comes from the $O(p^3)$ Lagrangian (\ref{piN3}).
{}From these Lagrangians one extracts the respective vertices (up to
trivial factors)
\beqa
k_2 &=& a_6 \tau_3 + \frac{a_7}{2} = \frac{1}{2}
\left( \ba{cc} 1 + \kappa_p & 0 \\ 0 & \kappa_n \ea \right) \nl
k_3 &=& \frac{1}{2} (1+ \tau_3) \left[ \left(a_6 - \frac{1}{8}\right)
\tau_3 + \frac{1}{2} \left( a_7 - \frac{1}{4}\right) \right] =
\frac{1}{4} \left( \ba{cc} 1 + 2\kappa_p & 0 \\ 0 & 0 \ea \right)
\eeqa
in terms of the nucleon anomalous magnetic moments $\kappa_N$.
The leading contribution to the spin--dependent Compton amplitude
in the forward direction for
small photon energies is of $O(p^3)$ and it is proportional to
\beq
k^2_2 - k_3 = \frac{1}{4} \left( \ba{cc} \kappa^2_p & 0 \\ 0 & \kappa^2_n
\ea \right)
\eeq
in accordance with the classic low--energy theorem
(Low, 1954; Gell-Mann and Goldberger, 1954)
\beq
c_1 = - \dfrac{\kappa_N^2 e^2}{8\pi m}~.
\eeq

%before each section
\setcounter{equation}{0}

\vspace*{.5cm}

\begin{center}
\section{CONCLUSIONS}
\label{sec:con}
\end{center}
CHPT is a systematic framework for analyzing the standard
model at low energies. It has all the desirable features of a quantum
field theory (unitarity, analyticity,\dots) even though it is
non--renormalizable. The transition from the fundamental level of
quarks and gluons to the effective level of hadrons generates a large
number of effective coupling constants (LECs). Since these constants
are not constrained by the symmetries of the standard model, additional
phenomenological and/or theoretical input is needed to make CHPT
predictive, especially in higher orders of the chiral expansion.

The chiral Lagrangian of the standard model is unique for the
chosen number $N_f$ of light flavours, but
\bit
\item it has many different parts as shown in Table 1, and
\item it can be organized in different ways (standard vs. generalized
CHPT).
\eit

Elastic pion--pion scattering provides an excellent example that
precise predictions are possible even to $O(p^6)$ (two--loop level)
although there is a forbidding number of 111 coupling constants in
the mesonic Lagrangian of $O(p^6)$. As chiral dimensional analysis
suggests, the corrections of $O(p^6)$ are indeed small. Moreover, they
are dominated by the unambiguous chiral logarithms whereas the
contribution of the LECs of $O(p^6)$
is very small, especially for the $S$--waves. Therefore, pion--pion
scattering is an almost ideal case for confronting QCD with
forthcoming precision experiments at low energies. At the level
of precision reached at $O(p^6)$, it is becoming necessary to include
electromagnetic and isospin violating corrections.

In the pion--nucleon or more generally the meson--baryon system,
we are still far from the precision attained in the meson sector.
There are several obvious reasons for this difference: the baryons
are not (pseudo--) Goldstone fields, resonances are in general
much closer to the physical threshold, the chiral expansion has
terms of all orders whereas only even orders appear in the meson
sector,\dots. Nevertheless, there is considerable progress both
on the more theoretical and on the phenomenological side also
in this field. Heavy baryon CHPT provides a systematic low--energy
expansion and the complete low--energy structure of the pion--nucleon
system is now known to $O(p^3)$. However, many things remain to
be done and the activity both in theory and experiment continues
to grow.

\vfill

\begin{center}
\section*{ACKNOWLEDGEMENTS}
\end{center}
I want to thank J. Bijnens, G. Colangelo, J. Gasser,
M. Moj\v zi\v s and M.E. Sainio for the enjoyable collaborations
that have led to the results presented here. It is also a
pleasure to thank A. F\"a\ss ler and the Ettore Majorana Centre
for Scientific Culture for the splendid hospitality in Erice.

\newpage

\begin{center}
\section*{REFERENCES}
\end{center}
\noindent
M. Baillargeon and P.J. Franzini (1995). In: Maiani et al.
(1995), p. 413.\\
V. Bernard, J. Gasser, N. Kaiser and U.-G. Mei\ss ner (1991). {\it Phys.
Lett.}, \ul{B268}, 291.\\
V. Bernard, N. Kaiser, J. Kambor and U.-G. Mei\ss ner (1992). {\it Nucl.
Phys.}, \ul{B388}, 315.\\
V. Bernard, N. Kaiser and U.-G. Mei\ss ner (1995). {\it Int. J. Mod.
Phys.}, \ul{E4}, 193.\\
A.M. Bernstein and B.R. Holstein, eds. (1995). {\it Chiral Dynamics:
Theory and Experiment}. Proc. of \\ \hspace*{.35cm} the Workshop
held at MIT, Cambridge,
MA, USA, July 1994. Springer, Berlin and Heidelberg.\\
J. Bijnens, G. Colangelo, G. Ecker, J. Gasser and M.E. Sainio (1995).
Elastic $\pi\pi$ scattering to \\ \hspace*{.35cm} two loops.
Preprint NORDITA-95/77 N,P, BUTP-95-34, UWThPh-1995-34,
HU-TFT-95-64; \\ \hspace*{.35cm} hep-ph/9511397.\\
C.G. Callan, S. Coleman, J. Wess and B. Zumino (1969). {\it Phys.
Rev.}, \ul{177}, 2247.\\
G. Colangelo (1995). {\it Phys. Lett.}, \ul{B350}, 85; {\it ibid.},
\ul{B361}, 234 (E).\\
S. Coleman, J. Wess and B. Zumino (1969). {\it Phys. Rev.},
\ul{177}, 2239.\\
G. Ecker, J. Gasser, A. Pich and E. de Rafael (1989a). {\it Nucl. Phys.},
\ul{B321}, 311.\\
G. Ecker, J. Gasser, H. Leutwyler, A. Pich and E. de Rafael (1989b).
{\it Phys. Lett.}, \ul{B223}, 425.\\
G. Ecker (1994). {\it Phys. Lett.}, \ul{B336}, 508.\\
G. Ecker (1995a). In: Bernstein and Holstein (1995), p. 41.\\
G. Ecker (1995b). {\it Progr. Part. Nucl. Phys.}, \ul{35}, 1
(A. F\"a\ss ler, ed.). Elsevier Science Ltd., Oxford.\\
G. Ecker and M. Moj\v zi\v s (1995c). Low--energy expansion of the
pion--nucleon Lagrangian. \\ \hspace*{.35cm}
{\it Phys. Lett. B} (in print); hep-ph/9508204.\\
H.W. Fearing and S. Scherer (1994). Extension of the chiral
perturbation theory meson Lagrangian  \\ \hspace*{.35cm} to order $p^6$.
Preprint TRI-PP-94-68; hep-ph/9408346.\\
J. Gasser and H. Leutwyler (1983). {\it Phys. Lett.}, \ul{125B}, 325.\\
J. Gasser and H. Leutwyler (1984). {\it Ann. Phys.}, \ul{158}, 142.\\
J. Gasser and H. Leutwyler (1985). {\it Nucl. Phys.}, \ul{B250}, 465.\\
J. Gasser, M.E. Sainio and A. \v Svarc (1988), {\it Nucl. Phys.}, \ul{B307},
779.\\
M. Gell-Mann and M.L. Goldberger (1954). {\it Phys. Rev.},
\ul{96}, 1428.\\
H. Georgi (1990). {\it Phys. Lett.}, \ul{B240}, 447.\\
B.R. Holstein (1990). {\it Phys. Lett.}, \ul{B244}, 83.\\
E. Jenkins and A.V. Manohar (1991). {\it Phys. Lett.}, \ul{B255}, 558.\\
E. Jenkins and A.V. Manohar (1992). In: Mei\ss ner (1992), p. 113.\\
M. Knecht and J. Stern (1995a). In: Maiani et al. (1995), p. 169.\\
M. Knecht, B. Moussallam and J. Stern (1995b). In: Maiani et al. (1995),
p. 221.\\
M. Knecht, B. Moussallam, J. Stern and N.H. Fuchs (1995c). The low energy
$\pi\pi$ amplitude to one \\ \hspace*{.35cm}
and two loops. Orsay preprint IPNO/TH 95-45; hep-ph/9507319.\\
A. Krause (1990). {\it Helvetica Phys. Acta}, \ul{63}, 3.\\
H. Leutwyler (1994a). {\it Ann. Phys.}, \ul{235}, 165.\\
H. Leutwyler (1994b). Principles of chiral perturbation theory.
Lectures given at the {\it Workshop  \\ \hspace*{.35cm} Hadron 94}, Gramado,
RS, Brasil. Univ. Bern preprint BUTP-94/13; hep-ph/9406283.\\
F. Low (1954). {\it Phys. Rev.}, \ul{96}, 1428;\\
M. Luke and A.V. Manohar (1992). {\it Phys. Lett.}, \ul{B286}, 348.\\
L. Maiani, G. Pancheri and N. Paver, eds. (1995). {\it The Second DA$\Phi$NE
Physics Handbook}.  \\ \hspace*{.35cm} INFN, Frascati.\\
T. Mannel, W. Roberts and Z. Ryzak (1992). {\it Nucl. Phys.},
\ul{B368}, 204.\\
U.-G. Mei\ss ner, ed. (1992). Proc. of the {\it Workshop on Effective
Field Theories of the Standard  \\ \hspace*{.35cm} Model}, Dobog\'ok\"o,
Hungary, 1991. World Scientific, Singapore.\\
M.M. Nagels et al. (1979). {\it Nucl. Phys.}, \ul{B147}, 189.\\
Y. Nambu and G. Jona-Lasinio (1961). {\it Phys. Rev.},
\ul{122}, 345; {\it ibid.}, \ul{124}, 246.\\
H. Neufeld and H. Rupertsberger (1995). {\it Z. Phys.}, \ul{C68}, 91.\\
A. Pich (1995). {\it Rep. Prog. Phys.}, \ul{58}, 563.\\
E. de Rafael (1995). In: {\it CP Violation  and the Limits of the Standard
Model}  \\ \hspace*{.35cm} (J.F. Donoghue, ed.). World Scientific,
Singapore. \\
Review of Particle Properties (L. Montanet et al.) (1994). {\it
Phys. Rev.}, \ul{D50}, 1173.\\
J. Stern, H. Sazdjian and N.H. Fuchs (1993). {\it Phys. Rev.}, \ul{D47},
3814.\\
R. Urech (1995). {\it Nucl. Phys.}, \ul{B433}, 234.\\
S. Weinberg (1966). {\it Phys. Rev. Lett.}, \ul{17}, 616.\\
S. Weinberg (1979). {\it Physica}, \ul{96A}, 327.\\
S. Weinberg (1990). {\it Phys. Lett.}, \ul{B251}, 288.\\
J. Wess and B. Zumino (1971). {\it Phys. Lett.}, \ul{37B}, 95.\\

\end{document}